\documentclass[twocolumn,showpacs,preprintnumbers,amsmath,amssymb]{revtex4}

\usepackage{graphicx}
\usepackage{dcolumn}
\usepackage{bm}

\newcommand{\bq}{\begin{equation}}
\newcommand{\eq}{\end{equation}}
\newcommand{\bqa}{\begin{eqnarray}}
\newcommand{\eqa}{\end{eqnarray}}
\newcommand{\nn}{\nonumber \\}

\def\be     {\begin{equation}}
\def\ee     {\end{equation}}
\def\bea        {\begin{eqnarray}}
\def\eea        {\end{eqnarray}}
\def\bnn    {\begin{eqnarray*}}
\def\enn    {\end{eqnarray*}}

\begin{document}

\title{Fermion zero mode and superfluid weight}
\author{Ki-Seok Kim}
\affiliation{Institute de Physique Th\'eorique, CEA, IPhT, CNRS,
URA 2306, F-91191 Gif-sur-Yvette, France}
\date{\today}

\begin{abstract}
We propose one possible mechanism for deconfinement based on an
SU(2) slave-boson theory. Resorting to an effective field theory
approach, we show that introduction of an isospin interaction
potential gives rise to a fermion zero mode in an
instanton-hedgehog configuration. As a result, meron-type vortices
are allowed. We demonstrate how emergence of such vortices results
in the doping-independent decreasing ratio of superfluid weight.
\end{abstract}


\maketitle

Study on strongly correlated electrons opened a new window so
called gauge theory in modern condensed matter physics beyond the
Fermi liquid theory and Landau-Ginzburg-Wilson framework of
"classical" condensed matter physics. When interactions are strong
enough compared with kinetic energy, an infinite interaction limit
can be a good starting point. The presence of such a large energy
scale gives rise to a constraint in dynamics of electrons. In
addition, interesting physics now arises from the kinetic-energy
contribution in the restricted Hilbert space, and "non-local"
order parameters or more carefully, link variables instead of
on-site ones in lattice models appear as important low energy
collective degrees of freedom. These link variables can be
formulated as gauge fields, and gauge theory arises naturally for
dynamics of strongly correlated electrons.

Slave-boson approach has been one of the canonical frameworks for
study of strongly correlated electrons. In particular, a doped
Mott insulator problem was formulated in the slave-boson
context,\cite{Lee_Nagaosa_Wen} where strong repulsive interactions
cause so called the single occupancy constraint naturally imposed
in the slave-boson representation, and link variables arise as
collective "order parameter" excitations formulated as gauge
fields. U(1) slave-boson gauge theory has been enjoyed both
intensively and extensively for the doped Mott insulator problem.

It will be not only fair but also true to say that this gauge
theoretical framework has explained many kinds of aspects
associated with high T$_{c}$ cuprates such as phase diagram,
thermodynamics, transport, spin dynamics, and
etc.\cite{Lee_Nagaosa_Wen} However, such a theoretical structure
seems to have fundamental difficulty for physics of high T$_{c}$
cuprates. One is about the presence of coherent electron-like
excitations near nodal directions in the normal state of high
T$_{c}$ cuprates.\cite{ARPES} Although the slave-boson framework
can recover Fermi liquid via condensation of a bosonic charge
degree of freedom (holon), such coherent electron excitations
disappear in an uncondensed phase as the above. This problem was
argued to be related with so called confinement in gauge
theory,\cite{Polyakov} one of the notorious problems in
theoretical physics, and it requires deeper understanding of
instanton physics in gauge fluctuations.

The other is associated with the doping-independent decreasing
ratio of superfluid weight.\cite{Larkin_SF} Although this problem
seems to be simple compared with the first one, information in
superfluid weight has important physical implication since it
reflects d-wave superconductivity emerging from a doped Mott
insulator. Unfortunately, the U(1) slave-boson framework cannot
explain this doping independence owing to charge renormalization
of nodal quasiparticles although it gives the correct zero
temperature superfluid density proportional to hole
concentration.\cite{Lee_Superfluid,DHLee_SF}

The second problem has motivated Wen and Lee to propose an SU(2)
slave-boson formulation, which extends the U(1) slave-boson theory
to include fluctuations between nearly degenerate U(1) mean-field
states, well applicable in underdoped
regions.\cite{Lee_Nagaosa_Wen} In the SU(2) slave-boson
representation an additional holon excitation called $b_{2}$ boson
appears to take such fluctuations into account. Based on this
formulation, they demonstrated how one can find the doping
independent decreasing ratio of superfluid weight. Assuming
confinement of spinons and holons into electrons, they could show
that such charge renormalization does not occur in the SU(2)
slave-boson formulation.\cite{Lee_Superfluid,Herbut_SF}

In this paper we revisit this issue based on one possible
deconfinement scenario. Our main observation is that the SU(2)
framework can give rise to an isospin interaction potential via
gauge fluctuations. Actually, such an isospin interaction was
introduced by Wen and Lee.\cite{NLsM} They studied its
perturbation effect, and found that Fermi segments of electron
excitations away from half filling can appear owing to the
presence of such a term although spinons are still at half filling
in the SU(2) slave-boson formulation.

In this paper we consider its non-perturbation effect. It is
important to notice that the SU(2) slave-boson gauge theory allows
meron-type vortices, and their tunnelling events from up meron
vortices to down ones are associated with instanton excitations of
U(1) gauge fields. In other words, a monopole configuration of
U(1) gauge fields induces a hedgehog pattern of holon isospins. As
a result, the effective field theory with an isospin interaction
term has a similar form with the SU(2) gauge theory well studied
by Jackiw and Rebbi, where 't Hooft monopole excitations are
allowed and a fermion zero mode exists due to the presence of an
isospin interaction.\cite{Jackiw} Actually, we find a fermion zero
mode from an explicit calculation in two space and one time
dimensions [$(2+1)D$]. Thus, instanton excitations are suppressed
and deconfinement of spinons and holons is realized. This scenario
was pursued in compact QED$_{3}$ without an isospin interaction by
Marston, but the zero mode was proven not to exist in such an
effective field theory.\cite{Marston}

Although our zero mode scenario is appealing, we should confess
that the fermion zero mode can become unstable because there is no
gap to protect the mode. We will discuss this possibility in more
detail.

Based on this zero-mode scenario to allow meron vortices, we
discuss superfluid weight. Resorting to a dual Lagrangian for
meron vortices, we find that charge renormalization does not occur
in a certain limit, thus resulting in the doping independent
decreasing ratio of superfluid weight.

Dynamics of doped holes in the antiferromagnetically correlated
spin background is described by the t-J Hamiltonian \bqa && H = -
t \sum_{\langle i j \rangle} (c_{i\sigma}^{\dagger}c_{j\sigma} +
H.c.) + J \sum_{\langle i j \rangle} (\vec{S}_{i}\cdot\vec{S}_{j}
- \frac{1}{4}n_{i}n_{j}) . \nn \eqa Introducing an SU(2)
slave-boson representation for an electron field \bqa &&
c_{i\uparrow} = \frac{1}{\sqrt{2}} h_{i}^{\dagger} \psi_{i+} =
\frac{1}{\sqrt{2}} (b_{i1}^{\dagger}f_{i1} +
b_{i2}^{\dagger}f_{i2}^{\dagger}) , \nn && c_{i\downarrow} =
\frac{1}{\sqrt{2}} h_{i}^{\dagger} \psi_{i-} = \frac{1}{\sqrt{2}}
(b_{i1}^{\dagger}f_{i2} - b_{i2}^{\dagger}f_{i1}^{\dagger}) , \eqa
where $\psi_{i+} = \left(
\begin{array}{c} f_{i1} \\ f_{i2}^{\dagger}
\end{array}\right)$ and $\psi_{i-} =  \left(
\begin{array}{c} f_{i2} \\ - f_{i1}^{\dagger}
\end{array}\right)$ are SU(2)
spinon-spinors  and $h_{i} =  \left(
\begin{array}{c} b_{i1} \\ b_{i2}
\end{array}\right)$ is holon-spinor, one can rewrite the t-J model
in terms of these fractionalized excitations with hopping and
pairing fluctuations \bqa && L = L_{0} + L_{s} + L_{h} , ~~~~~
L_{0} = J_{r} \sum_{\langle i j \rangle}
\mathbf{tr}[U_{ij}^{\dagger}U_{ij}]   , \nn && L_{s} = \frac{1}{2}
\sum_{i} \psi_{i\alpha}^{\dagger} (\partial_{\tau} - i
a_{i0}^{k}\tau_{k})\psi_{i\alpha} \nn && + J_{r} \sum_{\langle i j
\rangle} ( \psi_{i\alpha}^{\dagger}U_{ij}\psi_{j\alpha} + H.c.) ,
\nn && L_{h} =  \sum_{i} h_{i}^{\dagger}(\partial_{\tau} - \mu  -
i a_{i0}^{k}\tau_{k})h_{i} \nn && + t_{r} \sum_{\langle i j
\rangle} ( h_{i}^{\dagger}U_{ij}h_{j} + H.c.)  , \eqa where the
SU(2) matrix field is $U_{ij} =
\left( \begin{array}{cc} - \chi_{ij}^{\dagger} & \eta_{ij} \\
\eta_{ij}^{\dagger} & \chi_{ij} \end{array}\right)$, and $J_{r} =
\frac{3J}{16}$ and $t_{r} = \frac{t}{2}$ are redefined
couplings.\cite{Lee_Nagaosa_Wen} Since this decomposition
representation enlarges the original electron Hilbert space,
constraints are introduced via Lagrange multiplier fields
$a_{i0}^{k}$ with $k = 1, 2, 3$.

In the SU(2) formulation Wen and Lee choose the staggered flux
gauge\cite{Lee_Nagaosa_Wen} \bqa && U_{ij}^{SF} = - \sqrt{\chi^{2}
+ \eta^{2}} \tau_{3} \exp[i(-1)^{i_{x} + i_{y}} \Phi \tau_{3}]
\eqa with a phase $\Phi = \tan^{-1}\Bigl( \frac{\eta}{\chi}
\Bigr)$. Although the staggered flux ansatz breaks translational
invariance, this formal symmetry breaking is restored via SU(2)
fluctuations between nearly degenerate U(1) mean-field states. For
example, one possible U(1) ground state, the d-wave pairing one
$U_{ij}^{dSC} = - \chi \tau_{3} + (-1)^{i_{y} + j_{y}} \eta
\tau_{1}$ can result from the SU(2) rotation $U_{ij}^{dSC}= W_{i}
U_{ij}^{SF} W_{j}^{\dagger}$ with an SU(2) matrix $W_{i} =
\exp\Bigl\{ i(-1)^{i_{x} + i_{y}}\frac{\pi}{4} \tau_{1} \Bigr\}$.
Then, our starting point becomes the following effective
Lagrangian \bqa && L_{SF} = \frac{1}{2} \sum_{i}
\psi_{i\alpha}^{\dagger} (\partial_{\tau} - i
a_{i0}^{3}\tau_{3})\psi_{i\alpha} \nn && + J_{r} \sum_{\langle i j
\rangle} ( \psi_{i\alpha}^{\dagger}U_{ij}^{SF}
e^{ia_{ij}^{3}\tau_{3}}\psi_{j\alpha} + H.c.) \nn && + \sum_{i}
h_{i}^{\dagger}(\partial_{\tau} - \mu  - i
a_{i0}^{3}\tau_{3})h_{i} \nn && + t_{r} \sum_{\langle i j \rangle}
( h_{i}^{\dagger}U_{ij}^{SF} e^{ia_{ij}^{3}\tau_{3}}h_{j} + H.c.)
\nn && + J_{r} \sum_{\langle i j \rangle}
\mathbf{tr}[U_{ij}^{SF\dagger}U_{ij}^{SF}] , \eqa where we have
introduced only one kind of gauge field $a_{\mu}^{3}$ as important
low energy fluctuations since other two ones, $a_{\mu}^{1}$ and
$a_{\mu}^{2}$ are gapped due to Anderson-Higgs mechanism in the
staggered flux phase.

One can derive the following effective field theory from the
staggered flux ansatz of the SU(2) slave-boson theory Eq. (5),
\bqa && {\cal L}_{SF} = \bar{\psi}\gamma_{\mu}(\partial_{\mu} -
ia_{\mu}^{3}\tau_{3})\psi + \frac{1}{2g^{2}}
(\epsilon_{\mu\nu\gamma}\partial_{\nu}a_{\gamma}^{3})^{2} \nn && +
x \mathbf{z}^{\dagger}(\partial_{\tau} - ia_{0}^{3}\tau_{3} -
iA_{0})\mathbf{z} +
\frac{1}{u_{h}}|\mathbf{z}^{\dagger}(\partial_{\tau} -
ia_{0}^{3}\tau_{3} - iA_{0})\mathbf{z}|^{2} \nn && +
\frac{x}{2m_{b}}|(\partial_{i} - ia_{i}^{3}\tau_{3} -
iA_{i})\mathbf{z}|^{2} \nn && + x^{2}J \Bigl[
\frac{4}{c_1}|z_{1}|^{2}|z_{2}|^{2} + \frac{1}{c_2}(|z_{1}|^{2} -
|z_{2}|^{2})^{2} \Bigr]  . \eqa Dirac structure for spinon
dynamics\cite{Wen_Symmetry} results from the staggered flux gauge
in the SU(2) slave-boson theory, where $\psi$ is an $8$ component
spinor and Dirac gamma matrices are $\gamma_{0} = \left(
\begin{array}{cc} \sigma_{3} & 0
\\ 0 & - \sigma_{3} \end{array}\right)$, $\gamma_{1} = \left(
\begin{array}{cc} \sigma_{1} & 0 \\ 0 & - \sigma_{1}
\end{array}\right)$, and $\gamma_{2} = \left( \begin{array}{cc} \sigma_{2}
& 0 \\ 0 & - \sigma_{2} \end{array}\right)$. $a_{\mu}^{3}$ is the
remaining massless U(1) gauge field as an important low energy
degree of freedom in the staggered flux phase. We have introduced
a finite bare gauge charge $g$ in the long wave-length limit. It
is important to understand that spinons are still at half filling
even away from half filling in the SU(2) formulation. The
single-occupancy constraint in the SU(2) representation is given
by $f_{i1}^{\dagger}f_{i1} + f_{i2}^{\dagger}f_{i2} +
b_{i1}^{\dagger}b_{i1} - b_{i2}^{\dagger}b_{i2} = 1$. Thus, if the
condition of $\langle b_{i1}^{\dagger}b_{i1} \rangle = \langle
b_{i2}^{\dagger}b_{i2} \rangle = \frac{x}{2}$ with hole
concentration $x$ is satisfied, we see $\langle
f_{i1}^{\dagger}f_{i1} + f_{i2}^{\dagger}f_{i2} \rangle = 1$,
i.e., spinons are at half filling. As a result, a chemical
potential term does not arise in the spinon sector. Actually, this
was demonstrated for the staggered flux phase in the mean-field
analysis of the SU(2) slave-boson theory.\cite{NLsM}

Holon dynamics at low energies is described by CP$^{1}$ gauge
theory of the O(3) nonlinear $\sigma$ model\cite{Holon_Sector}
with a Berry phase contribution (the first term in the holon
sector) arising from finite density of holons $\mathbf{z} = \left(
\begin{array}{c} z_{+} \\ z_{-} \end{array}\right)$ while $z_{+}$
gauge charge is opposite to $z_{-}$ one. $u_{h}$ is associated
with compressibility for holons, and $m_{b} \sim 1/t$ is bare band
mass. The last two terms in the holon sector represent anisotropy
contributions of holon isospins, given by $\vec{I}_{hi} =
z_{i\alpha}^{\dagger}\vec{\tau}_{\alpha\beta}z_{i\beta} =
(\sin\theta_{i}\cos\phi_{i}, \sin\theta_{i}\sin\phi_{i},
\cos\theta_{i})$, where $c_{1}$ and $c_{2}$ are positive numerical
constants, possibly arising from short-distance fermion
fluctuations. In the SU(2) slave-boson theory each phase can be
identified with this isospin configuration, where the staggered
flux phase is characterized by $\langle \vec{I}_{h} \rangle = 0$
while the d-wave pairing state is expressed as $\langle I^{z}_{h}
\rangle = 0$ and $\langle {I}^{x(y)}_{h} \rangle \not= 0$. The
above effective field theory describes fluctuations between nearly
degenerate U(1) mean-field states via isospin fluctuations. The
last two terms favor an easy plane when $c_{1} > c_{2}$ expected
away from half filling.

In the easy plain limit one can represent the holon spinor as
$z_{\sigma} = e^{i\phi_{\sigma}}$. Then, the above effective field
theory becomes \bqa &&  {\cal L} =
\bar{\psi}_{\sigma}\gamma_{\mu}(\partial_{\mu} - i\sigma a_{\mu}
)\psi_{\sigma} \nn && +
\frac{1}{u_{h}}\Bigl(\partial_{\tau}\phi_{\sigma} - \sigma a_{0} -
A_{0} + i \frac{u_{h}}{2} x \Bigr)^{2} \nn && +
\frac{x}{2m_{b}}(\partial_{i}\phi_{\sigma} - \sigma a_{i} -
A_{i})^{2} + \frac{1}{2g^{2}}
(\epsilon_{\mu\nu\gamma}\partial_{\nu}a_{\gamma})^{2}  ,  \eqa
where $\pm$ are associated with the SU(2) space and the
superscript $3$ in the gauge field is omitted.

We first consider an infinite bare gauge coupling $g \rightarrow
\infty$, where U(1) gauge fields can be integrated out exactly
resulting in confinement.\cite{DHLee_SF,Nayak} Shifting gauge
fields as $a_{0} \rightarrow a_{0} - \partial_{\tau} \phi_{-} +
A_{0} - i \frac{u_{h}}{2} x$ and $a_{i} \rightarrow a_{i} -
\partial_{i}\phi_{-} + A_{i}$, and introducing field variables of
$c_{\sigma} = e^{i\sigma \phi_{-}}\psi_{\sigma}$ and $\phi_{c} =
\phi_{+} + \phi_{-}$, one can perform integration of U(1) gauge
fields and find an effective field theory for d-wave
superconductivity emerging from a doped Mott insulator in the
confinement limit \bqa && {\cal L} = \bar{c}_{\sigma} \gamma_{\mu}
\Bigl(\partial_{\mu} - \frac{i}{2}\sigma
\partial_{\mu}\phi_{c} \Bigr)c_{\sigma} \nn && + \frac{u_{h}}{8} (\sigma
\bar{c}_{\sigma}\gamma_{0}c_{\sigma})^{2} +
\frac{m_{b}}{4x}(\sigma \bar{c}_{\sigma}\gamma_{i}c_{\sigma})^{2}
\nn && + \frac{1}{2u_{h}}\Bigl(\partial_{\tau}\phi_{c} - 2A_{0} +
i u_{h}x \Bigr)^{2} + \frac{x}{4m_{b}}\Bigl(\partial_{i}\phi_{c} -
2 A_{i} \Bigr)^{2} . \eqa It should be noted that all dynamic
variables are gauge singlets, thus gauge invariance associated
with $a_{\mu}$ is satisfied automatically. As shown in this field
theory, charge renormalization does not occur although the phase
stiffness is proportional to hole concentration implying that this
superconductivity results from a doped Mott insulator. This is in
contrast with the U(1) formulation, where electric charge of
quasiparticles is proportional to hole
concentration.\cite{Lee_Superfluid,DHLee_SF} As a result, we
obtain the following expression for superfluid weight
$\rho_{s}(x,T) = \rho_{s}(x) - c T$, where the decreasing ratio
$\frac{d \rho_{s}(x,T)}{d T} = - c$ is doping independent.
Actually, this confinement scenario was investigated previously,
but in a different version.\cite{Lee_Superfluid}

If the bare gauge charge is finite, one cannot perform integration
of gauge fluctuations exactly. Considering that spinon
contributions give rise to the following renormalized gauge
dynamics $\frac{N}{16}(\partial\times{a})
\frac{1}{\sqrt{-\partial^{2}}}(\partial\times{a})$ with the flavor
number $N$, and performing duality transformation for holon
vortices and instantons, we find an effective Lagrangian \bqa &&
{\cal L}_{dual} = |(\partial_{\mu} -
ic_{\mu}^{\sigma})\Phi_{\sigma}|^{2} +
m_{v}^{2}|\Phi_{\sigma}|^{2} + V_{eff}(|\Phi_{\sigma}|) \nn && +
\frac{u_{h}}{4}[(\partial\times c_{\sigma})_{\tau} - x]^{2} +
\frac{m_{b}}{2x}(\partial\times c_{\sigma})_{i}^{2} \nn && +
\frac{4}{N}(\partial_{\mu}\varphi - c_{+} +
c_{-})\sqrt{-\partial^{2}}(\partial_{\mu}\varphi - c_{+} + c_{-})
\nn && - y_{m}(e^{i\varphi}\Phi_{+}^{\dagger}\Phi_{-} + H.c.) .
\eqa $\Phi_{\pm}$ represent holon vortices, and $c_{\mu}^{\pm}$
are vortex gauge fields associated with their long-range
interactions. $\varphi$ is a magnetic potential field associated
with instanton excitations of U(1) gauge fields $a_{\mu}$, and
$y_{m}$ is an instanton fugacity. In particular, the last term
describes tunnelling events between $+$ and $-$ meron vortices,
where instanton excitations are summed in the dilute
approximation.

Eq. (9) has been studied intensively in the context of quantum
antiferromagnetism,\cite{DQCP1,DQCP2} where fermion excitations
are gapped thus ignored in the low energy limit, allowing a
simpler expression ${\cal L}_{dual} = |(\partial_{\mu} -
ic_{\mu})\Phi_{\sigma}|^{2} + m_{v}^{2}|\Phi_{\sigma}|^{2} +
V_{eff}(|\Phi_{\sigma}|) + \frac{u_{h}}{2}[(\partial\times
c)_{\tau} - x]^{2} + \frac{m_{b}}{x}(\partial\times c)_{i}^{2} -
y_{m}([\Phi_{+}^{\dagger}\Phi_{-}]^{n} + H.c.)$ with $x = 0$ and
integer $n$. Here, the additional multiplication with $n$ is
argued to arise from a Berry phase contribution of gauge field.
Although gapless fermion excitations renormalize gauge dynamics,
the instanton-induced term will be relevant in both staggered flux
and superconducting phases, i.e., away from quantum criticality.
On the other hand, such instanton fluctuations are claimed to be
irrelevant at the quantum critical point even in the $n=1$ case,
allowing meron-type vortices at the quantum critical
point.\cite{DQCP2}

We propose a deconfinement mechanism which has nothing to do with
critical fluctuations\cite{DQCP1,DQCP2,DQCP3}. Mentioned in the
introduction, a fermion zero mode can emerge to suppress instanton
effects if the effective field theory Eq. (6) is modified
slightly. It has been argued that gauge fluctuations of time
components can induce an isospin interaction potential.\cite{NLsM}
Introducing an isospin interaction term ${\cal L}_{I} = U_{I}
\bar{\psi} (\vec{I}_{h}\cdot\vec{\tau})\psi$ with its coupling
strength $U_{I}$, we obtain an equation of motion for Dirac
fermions in $(2+1)D$ \bqa && \gamma_{\mu}(\partial_{\mu} -
ia_{3\mu}^{cl} \tau_{3})\psi + U_{I}
(\vec{I}_{h}^{cl}\cdot\vec{\tau})\psi = E \psi , \eqa where $E$ is
an eigen value.

A full procedure should be as follows. Starting from Eq. (6) with
the isospin coupling term, we derive two equations of motion for
holons and spinons, respectively. Considering an instanton
configuration in the gauge potential, we solve the holon sector
and find its corresponding holon configuration. Remember that the
holon sector is exactly the same as the CP$^{1}$ gauge theory if
the contribution from finite density of holons is neglected, i.e.,
the linear time-derivative term. The $\sigma$ model study has
shown that an instanton potential gives rise to a hedgehog
configuration of spins in $(2+1)D$.\cite{Sachdev_Skyrmion}
Actually, this contribution is reflected in the instanton-induced
hedgehog term of the dual vortex Lagrangian, i.e., $- y_{m}
(\Phi_{+}^{\dagger}\Phi_{-} + H.c.)$. In the present paper we do
not solve the holon sector and assume the presence of such a
solution. As far as hole concentration is not too large, we expect
that our zero mode scenario may be applicable at least since the
contribution from finite density of holons (the Berry phase term
in the holon sector) would not spoil such a
configuration.\cite{Hedgehog_NLsM}

Based on this discussion, we solve the fermion part Eq. (10). The
presence of an isospin interaction term reminds us of the SU(2)
gauge theory in terms of massless Dirac fermions and adjoint Higgs
fields interacting via SU(2) gauge fields, where topologically
nontrivial stable excitations called 't Hooft-Polyakov monopoles
are allowed. Jackiw and Rebbi have shown that the Dirac equation
with isospin couplings has a fermion zero mode in a 't
Hooft-Polyakov monopole potential.\cite{Jackiw} Actually, we see
that such a fermion zero mode exists indeed in our effective field
theory. Existence of a fermion zero mode allows deconfinement of
spinons and holons.

Considering an instanton configuration $a_{3\mu}^{cl} = a(s)
\epsilon_{3\nu\mu}x_{\nu}$ with $a(s) \sim \frac{1}{s^{2}}$ for $s
\rightarrow \infty$ where $s = \sqrt{\tau^{2} + x^{2} + y^{2}}$
and its corresponding isospin hedgehog configuration
$I_{h\mu}^{cl} = \Phi(s) x_{\mu}$ with $\Phi(s) \sim \frac{1}{s}$
for $s \rightarrow \infty$, and inserting the $4$ component spinor
$\psi_{n} = \left( \begin{array}{c} \chi_{n}^{+} \\ \chi_{n}^{-}
\end{array}\right)$ with an isospin index $n = 1, 2$ into Eq. (10),
where $\chi_{n}^{\pm}$ represent $2$ component spinors, we obtain
\bqa && \sigma_{ij}^{3}
\partial_{\tau}\chi^{\pm}_{jn} + \sigma_{ij}^{1}\partial_{x}
\chi^{\pm}_{jn} + \sigma^{2}_{ij}\partial_{y} \chi^{\pm}_{jn} \nn
&& + ia(s)y \sigma^{1}_{ij}\chi^{\pm}_{jm}\tau_{mn}^{3} - ia(s)x
\sigma^{2}_{ij}\chi^{\pm}_{jm}\tau_{mn}^{3} \nn && \pm U_{I}
\Phi(s) \chi^{\pm}_{im} (x_{\mu}\tau_{\mu}^{T})_{mn} = 0 . \eqa
Here, the zero mode condition $E = 0$ is utilized. Observing the
fact that discrimination between the isospin space and Dirac one
disappears in the above expression, one can replace the
$\vec{\tau}$ matrix with $\vec{\sigma}$.

Inserting the following expression $\chi^{\pm}_{jn} =
\mathcal{M}^{\pm}_{jm}\sigma^{3}_{mn}$ with
$\mathcal{M}^{\pm}_{jm} = g^{\pm}\delta_{jm} +
g_{\mu}^{\pm}\sigma^{\mu}_{jm}$ into the above, where any $2
\times 2$ matrix $\mathcal{M}^{\pm}_{jm}$ can be represented with
unit and Pauli matrices, we find \bqa && [\partial_{\tau} \pm
U_{I}\Phi(s) \tau] g^{\pm} + i [\partial_{x} - a(s) x \pm
U_{I}\Phi(s) x]g_{2}^{\pm} \nn && - i [\partial_{y} - a(s) y \mp
U_{I}\Phi(s) y]g_{1}^{\pm} = 0 , \nn && - [\partial_{\tau} \mp
U_{I}\Phi(s) \tau] g_{1}^{\pm} + [\partial_{x} + a(s) x \pm
U_{I}\Phi(s) x]g_{3}^{\pm} \nn && + i [\partial_{y} + a(s)y \pm
U_{I}\Phi(s) y]g^{\pm} = 0 , \nn && - [\partial_{\tau} \mp
U_{I}\Phi(s) \tau] g_{2}^{\pm} - i [\partial_{x} + a(s) x \mp
U_{I}\Phi(s) x]g^{\pm} \nn && + [\partial_{y} + a(s) y \mp
U_{I}\Phi(s) y ]g_{3}^{\pm} = 0 , \nn && [\partial_{\tau} \pm
U_{I}\Phi(s) \tau] g_{3}^{\pm} + [\partial_{x} - a(s) x \mp
U_{I}\Phi(s) x]g_{1}^{\pm} \nn && + [\partial_{y} - a(s) y \pm
U_{I}\Phi(s) y]g_{2}^{\pm} = 0 . \eqa As a result, we find a zero
mode equation \bqa && [\partial_{\tau} + U_{I}\Phi(s) \tau]
g_{1}^{-}  = 0 , \nn && [\partial_{x} - a(s) x + U_{I}\Phi(s)
x]g_{1}^{-} = 0 , \nn && [\partial_{y} - a(s) y + U_{I}\Phi(s)
y]g_{1}^{-} = 0 , \eqa yielding one normalizable fermion zero mode
$g_{1}^{-} \sim \exp\Bigl[-\int{d\tau}U_{I}\Phi(s)\tau\Bigr]
\exp\Bigl[\int{d\vec{r}}\cdot\vec{r}\Bigl(a(s) -
U_{I}\Phi(s)\Bigr) \Bigr]$. We note that this solution is
basically the same as that of the SU(2) gauge theory by Jackiw and
Rebbi.\cite{Jackiw}

More fundamentally, the presence of such a fermion zero mode
coincides with an index theorem,\cite{Polyakov} stating that
difference between the number of a zero mode with a left chirality
and that with a right chirality is the same as a topological
charge of the vacuum state. Since the topological vacuum charge is
one, only one zero mode with a left chirality $-$ is found indeed.

Integrating out Dirac fermions, we see the partition function of
instanton contributions as follows $Z_{M} = \exp\Bigl[ N
\mathbf{tr} \ln \Bigl\{\gamma_{\mu} (
\partial_{\mu} - ia^{cl}_{3\mu}\tau_{3} ) +
U_{I} (\vec{I}_{h}^{cl}\cdot\vec{\tau}) \Bigr\} \Bigr]$. Since the
eigen value of the argument matrix is zero, the partition function
vanishes, implying that single instanton excitations will be
suppressed. This serves one possible deconfinement
mechanism\cite{Kim_ZM} that has nothing to do with critical
fluctuations.

The final expression of our effective field theory becomes in the
easy plane limit $c_{1} > c_{2}$ \bqa && {\cal L}_{ZM} =
\bar{\psi}_{\sigma}\gamma_{\mu}(\partial_{\mu} - i\sigma a_{\mu}
)\psi_{\sigma} + U_{I} \bar{\psi} (\vec{I}_{h}\cdot\vec{\tau})\psi
\nn && + \frac{1}{u_{h}}\Bigl(\partial_{\tau}\phi_{\sigma} -
\sigma a_{0} - A_{0} + i \frac{u_{h}}{2} x \Bigr)^{2} \nn && +
\frac{x}{2m_{b}}(\partial_{i}\phi_{\sigma} - \sigma a_{i} -
A_{i})^{2} +
\frac{1}{2g^{2}}(\epsilon_{\mu\nu\gamma}\partial_{\nu}a_{\gamma})^{2}
, \eqa where U(1) gauge field is now noncompact. $\pm$ represent
an isospin index. Since a non-perturbation effect of the isospin
coupling term is already introduced in this expression, it will
give a perturbation contribution such as a chemical potential term
for nodal quasiparticles in the normal state.

Performing the duality transformation where perturbation effects
of the isospin term are ignored, we find the vortex Lagrangian
\bqa && {\cal L}_{dual} = |(\partial_{\mu} -
ic_{\mu}^{\sigma})\Phi_{\sigma}|^{2} +
m_{v}^{2}|\Phi_{\sigma}|^{2} + V_{eff}(|\Phi_{\sigma}|) \nn && +
\frac{u_{h}}{4}[(\partial\times c_{\sigma})_{\tau} - x]^{2} +
\frac{m_{b}}{2x}(\partial\times c_{\sigma})_{i}^{2} \nn && - i
(\epsilon_{\mu\nu\lambda}\partial_{\nu}a_{\lambda}) ( c_{\mu}^{+}
- c_{\mu}^{-}) - i
(\epsilon_{\mu\nu\lambda}\partial_{\nu}A_{\lambda}) ( c_{\mu}^{+}
+ c_{\mu}^{-}) \nn && +
\bar{\psi}_{\sigma}\gamma_{\mu}(\partial_{\mu} - i\sigma a_{\mu}
)\psi_{\sigma} + \frac{1}{2g^{2}}
(\epsilon_{\mu\nu\lambda}\partial_{\nu}a_{\lambda})^{2} . \eqa
Notice that the vortex-tunnelling term does not appear. As a
result, meron vortices are allowed.

To discuss superfluid weight, we consider the superconducting
phase characterized by $\langle \Phi_{\sigma} \rangle = 0$. Then,
Eq. (15) reads \bqa && {\cal L}_{eff} =
\frac{1}{2\rho_{\sigma}}(\epsilon_{\mu\nu\lambda}\partial_{\nu}c_{\lambda}^{\sigma})^{2}
- i (\epsilon_{\mu\nu\lambda}\partial_{\nu}a_{\lambda}) (
c_{\mu}^{+} - c_{\mu}^{-}) \nn && - i
(\epsilon_{\mu\nu\lambda}\partial_{\nu}A_{\lambda}) ( c_{\mu}^{+}
+ c_{\mu}^{-}) + \bar{\psi}_{\sigma}\gamma_{\mu}(\partial_{\mu} -
i\sigma a_{\mu} )\psi_{\sigma} \nn && + \frac{1}{2g^{2}}
(\epsilon_{\mu\nu\lambda}\partial_{\nu}a_{\lambda})^{2} , \eqa
where $\rho_{\pm}$ are stiffness parameters for $\pm$ holon
fields, respectively.

Integrating out both vortex gauge fields $c_{\mu}^{\pm}$ and
slave-boson U(1) gauge fields $a_{\mu}$, we find an effective
Lagrangian \bqa && {\cal L}_{eff} =
\bar{\psi}_{\sigma}\gamma_{\mu}\Bigl\{\partial_{\mu} - i
\Bigl(\frac{\rho_{+} - \rho_{-}}{\rho_{+} + \rho_{-}}\Bigr)\sigma
A_{\mu} \Bigr\} \psi_{\sigma} \nn && + \frac{1}{2(\rho_{+} +
\rho_{-})}(\sigma \bar{\psi}_{\sigma}\gamma_{\mu}
\psi_{\sigma})^{2} + \frac{2\rho_{+}\rho_{-}}{\rho_{+} + \rho_{-}}
A_{\mu}^{2} . \eqa This expression is quite interesting since
charge renormalization of nodal quasiparticles does not occur if
we assume $\rho_{+} >> \rho_{-}$. Actually, this ansatz seems to
be reasonable since $b_{+}$ holons exist in the U(1) formulation
but $b_{-}$ holons do not, reflecting reduction of the SU(2)
symmetry down to U(1) away from half filling. In this limit we
find the effective field theory ${\cal L}_{eff} =
\bar{\psi}_{\sigma}\gamma_{\mu} (\partial_{\mu} - i \sigma A_{\mu}
) \psi_{\sigma} + \frac{1}{2 \rho_{+} }(\sigma
\bar{\psi}_{\sigma}\gamma_{\mu} \psi_{\sigma})^{2} +  2 \rho_{-}
A_{\mu}^{2}$. An additional condition $\rho_{-} \propto x$ is
necessary in order to obtain $T_{c} \propto x$ from the superfluid
formulation. Despite this unsatisfactory point, we can see now why
the superconducting transition temperature is not so high compared
with the prediction of the U(1) slave-boson
theory.\cite{Lee_Superfluid} We note $T_{c} \propto \rho_{+}$ in
the U(1) formulation, but $T_{c} \propto \rho_{-}$ in the SU(2)
one. $\rho_{+} >> \rho_{-}$, i.e., the contribution from SU(2)
fluctuation explains this.

Several remarks are in order. First, one cautious theorist may
suspect stability of the zero mode solution since there is
expected to be no gap for spinons that remain at half-filling. In
this respect the present situation for existence of a fermion zero
mode is different from that discussed in Ref. \cite{Kim_ZM} for
antiferromagnetism, where the zero mode is a mid-gap state due to
the presence of antiferromagnetism. At present, we cannot exclude
that such a "gapless" zero mode may be unstable via quantum
fluctuations. In this case the fermion-zero-mode mechanism for
deconfinement is not applicable, and another mechanism associated
with quantum criticality may be available.\cite{DQCP1,DQCP2,DQCP3}
Second, one may consider that the ad hoc limit $\rho_{+}
>> \rho_{-}$ introduced for explanation of the temperature
dependence of the superfluid density is not consistent with the
condition for existence of a fermion zero mode, that is, $\langle
b_{i1}^{\dagger}b_{i1} \rangle = \langle b_{i2}^{\dagger}b_{i2}
\rangle = \frac{x}{2}$ with hole concentration $x$, satisfied in
the staggered flux phase. An important point is that $\rho_{+}$
and $\rho_{-}$ are not the same as the density of each boson,
respectively. As far as we know, this problem of interacting
bosons with gauge fluctuations is not fully understood yet. Each
holon density will be different from each superfluid density, thus
the condition of $\rho_{+} >> \rho_{-}$ should be regarded as
another one, not inconsistent with the zero mode scenario. Third,
one can propose a fermion zero mode localized in the meron vortex
configuration instead of the hedgehog configuration. Actually,
this is possible. Then, such meron vortices can acquire fermionic
quantum numbers, for example spin.\cite{Jackiw} Furthermore, their
statistics may turn into fermions.\cite{Abanov_WZW} This rich
possibility may open an important direction for study of high
T$_{c}$ cuprates.

One may ask whether the present zero mode scenario is consistent
with the recent scanning tunnelling microscopy (STM)
data\cite{STM} indicating columnar modulation in local density of
states (LDOS). Unfortunately, we cannot say anything about the
structure of vortices since vortices are taken into account as
point particles in our effective field theory approach. What we
can say is that tunnelling events between meron vortices can be
suppressed via the fermion zero mode, implying that the SU(2)
meron vortex will have the staggered flux core.\cite{SF_Core}
Actually, LDOS in the staggered-flux vortex core has been
discussed in Ref. \cite{STM_Theory}, but a direct comparison with
the recent data is not clear. One possible scenario is as follows.
In the staggered vortex core a chemical potential term arises for
Dirac fermions since the density of $b_{1}$ bosons is different
from that of $b_{2}$ ones. Then, hole pockets are allowed inside
the vortex core. In this situation one may argue that
quasiparticle scattering events near Dirac nodes can give rise to
such density modulation, intensively discussed
before.\cite{CDW_QPSC} This argument differs from the Cooper pair
density wave as a possible Berry phase effect.\cite{Tesanovic}

In this paper we proposed one mechanism how meron-type vortices
can appear beyond confinement based on a field theory approach,
where such vortices are taken to be point particles although they
have complex structures at short wave length
scales.\cite{Lee_Nagaosa_Wen} We have seen that a fermion zero
mode emerges owing to a special structure of the SU(2) slave-boson
theory. Furthermore, we demonstrated how the presence of such
vortices can allow the doping independent decreasing ratio of
superfluid weight.

This work is supported by the French National Grant ANR36ECCEZZZ.
K.-S. Kim is also supported by the Korea Research Foundation Grant
(KRF-2007-357-C00021) funded by the Korean Government.


\begin{thebibliography}{9}
\bibitem{Holon_Sector} Inserting the following expression $h_{i}^{SF} = (\sqrt{x} +
\delta b_{i}) \left(
\begin{array}{c} z_{i1} \\ - i (-1)^{i_{x}+i_{y}} z_{i2}
\end{array}\right)$ with $z_{i1} = e^{i\alpha_{i}} e^{-i\frac{\phi_{i}}{2}}
\cos\frac{\theta_{i}}{2}$ and $z_{i2} = e^{i\alpha_{i}}
e^{i\frac{\phi_{i}}{2}} \sin\frac{\theta_{i}}{2}$ into the holon
sector of the SU(2) slave-boson theory Eq. (5), where this
representation can be derived easily from an SU(2) rotation of
$h_{i}^{SF} = g_{i}^{\dagger}h_{i}^{dSC}$ with an SU(2) matrix
field $g_{i}^{\dagger} = \exp\Bigl[ - i (-1)^{i_{x}+i_{y}}
\frac{\theta_{i}}{2} \tau_{1} \Bigr] \exp\Bigl[ - i
\frac{\phi_{i}}{2} \tau_{3} \Bigr]$ and $h_{i}^{dSC} = \left(
\begin{array}{c} \sqrt{x} \\ 0 \end{array}\right)$, and performing
integration of amplitude fluctuations $\delta b_{i}$, we find an
effective field theory for the holon sector in Eq. (6).\cite{NLsM}
\bibitem{Hedgehog_NLsM} This argument can be further supported by
the dual vortex Lagrangian Eq. (9), where the instanton-induced
hedgehog term is allowed in the presence of an effectively applied
static magnetic field $\frac{u_{h}}{4}[(\partial\times
c_{\sigma})_{\tau} - x]^{2}$, resulting from the finite density
$x$ of holons. Although interplay between the presence of finite
magnetic fields and the instanton term is not known, it is
expected to be negligible as far as hole concentration is small.
\bibitem{SF_Core} Note that this staggered flux core is in
U(1), different from the staggered flux phase mentioned in the
SU(2) slave-boson theory, where the former is represented by
$\langle I_{h}^{z} \rangle \not= 0$ with $\langle I_{h}^{x(y)}
\rangle = 0$ while the latter is expressed as $\langle \vec{I}_{h}
\rangle = 0$.
\end{thebibliography}
\end{document}